# Are Scopus journal field classifications ever misleading?

Mike Thelwall, Information School, University of Sheffield, Sheffield, UK.
https://orcid.org/0000-0001-6065-205X m.a.thelwall@sheffield.ac.uk
Stephen Pinfield, Information School, University of Sheffield, Sheffield, UK.
https://orcid.org/0000-0003-4696-764X s.pinfield@sheffield.ac.uk

Journal field classifications in Scopus are used for citation-based indicators and by authors choosing appropriate journals to submit to. Whilst prior research has found that Scopus categories are occasionally misleading, it is not known how this varies for different journal types. In response, we assessed whether specialist, cross-field and general academic journals sometimes have publication practices that do not match their Scopus classifications. For this, we compared the Scopus narrow fields of journals with the fields that best fit their articles' titles and abstracts. We also conducted qualitative follow-up to distinguish between Scopus classification errors and misleading journal aims. The results show sharp field differences in the extent to which both cross-field and apparently specialist journals publish articles that match their Scopus narrow fields, and the same for general journals. The results also suggest that a few journals have titles and aims that do not match their contents well, and that some large topics spread themselves across many relevant disciplines. Thus, the likelihood that a journal's Scopus narrow fields reflect its contents varies substantially by field (although without systematic field trends) and some cross-field topics seem to cause difficulties in appropriately classifying relevant journals. These issues undermine citation-based indicators that rely on journal-level classification and may confuse scholars seeking publishing venues.

# Introduction

Scientific knowledge is primarily reported through academic journal articles, with scholars conventionally relying on journals to act as gatekeepers for academic research. Most journals not only filter for quality but also for topic, flagging this with their titles and aims. For example, the apparent scope of the Journal of Computational and Applied Mathematics can be guessed from its title, and this is narrowed down for the specialist in its Aims & Scope statements like, "The main interest of the Journal is in papers that describe and analyse new computational techniques for solving scientific or engineering problems. [] The computational efficiency (e.g., the convergence, stability, accuracy, ...) should be proved and illustrated by nontrivial numerical examples."[1] This information helps new authors seeking appropriate publication venues and readers seeking to browse relevant journals. It is also used by owners of scholarly article indexes, such as the Web of Science and Scopus, to classify each journal into one or more academic fields. These classifications are then used in turn to create citation-based impact indicators, such as field normalised citation counts for articles, field-based journal league tables, and field normalised journal impact calculations. For all these uses, it is important to assess the extent to which journals are given appropriate subject classifications in major databases and the extent to which a journal's title and aims are reliable indicators of its contents.

---

[1] https://www.sciencedirect.com/journal/journal-of-computational-and-applied-mathematics

It is useful, in theory, to distinguish between journal types because journals vary in specificity and therefore the extent to which they can be expected to match any given field (e.g., Boyack & Klavans, 2011). From the perspective of an individual field, we have identified four types of journals: first, specialist journals that primarily aim to publish articles from the field; second, cross-field (or generalist) journals that aim to publish articles in that field and some others; third, general journals that aim to publish in all areas of science; and fourth, out-of-field journals that do not aim to publish articles in the field (whether or not they actually publish articles within that field). For the last class, we are primarily interested in out-of-field journals that frequently publish articles within the field, for whatever reason, as described below. Cross-field journals that are not fully general can vary in breadth from two clear disciplines (e.g., education and sociology) to a broad remit (e.g., all life sciences) and some address cross-disciplinary issues (e.g., social science research methods). A journal's aims may not match its Scopus narrow field(s) or its publishing practice, however (Table 1). This article investigates the extent to which journal Scopus classifications match publishing practices (i.e., the contents of the articles published) for different journal types.

Table 1. A categorisation of journal types for the example of the Library and Information Science (LIS) field. From a LIS perspective, every journal is of exactly one type from each of the three classifications (columns 2-4).

| Journal type | Journal aim | Scopus narrow field(s) of journal | Journal publishing practice | Examples |
|---|---|---|---|---|
| Specialist (LIS) | Publish LIS articles only (all field or specialism within field, e.g., libraries). | Library and Information Sciences only. | LIS-related articles to a much greater extent than other fields. | *Serials Librarian* has LIS aims, is solely in the LIS Scopus narrow field, with a (subset of) LIS scope. |
| Cross-field (LIS) | Publish in LIS and one or more other fields. | Library and Information Sciences and one or more other narrow field(s). | LIS-related articles to a similar extent to a few other fields. | *Journal of Education for Library and Information Science* has cross-field aims and two corresponding Scopus narrow fields (LIS and Education), as reflected in its articles. |
| General | Publish in all fields. | Multidisciplinary | Similar coverage of all fields, including LIS. | *Science, Nature* have science wide aims and are in the Scopus Multidisciplinary narrow field. They publish occasional LIS-related pieces, but probably few compared to genetics. |
| Out-of-field (non-LIS) | Primarily publish in some non-LIS field(s). | Not Library and Information Sciences; not Multidisciplinary. | Primarily publishes articles in one or a few non-LIS field(s). | *Research Policy* has arguably non-LIS aims and is not in the Scopus LIS narrow field. Although it publishes some LIS-related articles, they are probably rare. |

One previous study has assessed the extent to which journals are classified correctly across science, comparing Scopus and the Web of Science 2010-14. It used direct citations only, identifying journals as potentially unrelated to a field that they had been assigned to if a low proportion of their direct citations were within that field (i.e., to or from other journals in the same field). It identified journals as potentially relevant to a field that they had not been assigned to if a high proportion of their direct citations started or finished in that field. There were 32 Scopus journals failing both criteria, usually because they had a misleading title, or their scope statement did not match their contents (Wang & Waltman, 2016). The aims of this study were like those of the current paper (and more ambitious in the sense that it investigated two databases) but the direct citation method was not comprehensive because all journals had to have at least 100 direct citations to be included, which 24% did not, and these were mainly new journals or from the arts and humanities. The current paper updates this previous study by 8 years (during which the classification errors previously identified may have been corrected) but uses text matching instead of citations to give a different perspective and potentially greater arts and humanities coverage (although this did not occur), and separates cross-field from specialist and out-of-field journals.

In addition to the above large-scale study, journal classification has also been investigated for information science and library science with direct citations, finding inconsistencies (Leydesdorff & Bornmann, 2016). Many studies have also proposed algorithms to improve the classification of journals, including for Scopus (Gómez-Núñez et al., 2016), but there does not seem to be a clear favourite. A different approach has been to develop alternative classification schemes (Archambault et al., 2011; Börner et al., 2012).

The term field here is used in the broad and loose sense of a body of related academic research. This contrasts with a discipline, which is a field supported by additional structures, such as journals, departments, and professional organisations (Sugimoto & Weingart, 2015) as well as established cultures (Becher & Trowler, 2001; Trowler et al., 2012). The Scopus narrow fields seem to roughly correspond to disciplines and so both terms will be used here, for linguistic variety. Journals may be created to support communities of scholars (e.g., Urbano et al., 2020), suggesting a tendency to increased specialisation, but there is also an opposite trend for publishing in general open access megajournals (Spezi et al., 2017). In between these two, a journal may represent an interdisciplinary field that spans the boundaries of traditional fields but is still relatively narrow in scope. An example of this is the journal *Scientometrics*, which incorporates elements of library and information science, computing, science and technology studies, and research policy, whilst retaining a narrow publishing practice.

As mentioned above, the current study, although with different goals, partly updates a previous analysis of the extent to which journals match their database categories (Wang & Waltman, 2016), tests a new text comparison method, and differentiates between journal types in the analyses. It also takes a different perspective on reporting the results by focusing on field/disciplinary differences and extreme cases rather than reporting the number of journals that exceed pre-defined thresholds. The following questions drive the study.

- **RQ1**: Are there field differences in the extent to which **specialist** Scopus journals publish articles that match their Scopus field?
- **RQ2**: Are there field differences in the extent to which **cross-field** Scopus journals publish articles that match their Scopus fields?
- **RQ3**: Do **general** Scopus journals (i.e., classified as Multidisciplinary) ever have more specialist publishing practices?
- **RQ4**: Are there field differences in the type (specialist, specialist and cross-field) of **out-of-field** journals?
- **RQ5**: Why do some journals mainly publish articles not matching their Scopus narrow fields?

# Method

The first stage (RQ1- RQ4) of the overall research design was to gather a large science-wide sample of academic journals for the most recent complete year, 2022, and compare the journals' Scopus narrow field(s) with the estimated Scopus narrow field(s) of the articles published in the journals (i.e., the journals' publishing practices as defined in Table 1), seeking discrepancies. The second stage (RQ5) was to qualitatively analyse the results to make inferences about the match between journal Scopus narrow field(s) and journal publishing practices. Scopus was chosen as the source of the journals since it has wider coverage than the Web of Science (Martín-Martín et al., 2021) and, unlike Dimensions.ai and Google Scholar, it has an established manual fine-grained field classification system for journals that was needed in a subsequent step. Brief software instructions are in the Supplementary file.

**Strategy for identifying journal type in Scopus**: Although there are too many journals (26,233 in Scopus for 2022) to manually classify their aims into fields, given that they are expressed in different terminologies, in different places, with specialist terminology, the existing Scopus journal classifications are straightforward to translate into journal types in Scopus. Journal Scopus narrow field(s) are assigned every year by an interdisciplinary team. Presumably, given the number of journals to consider, they are mainly applied to new journals. It seems likely that the classification team would consider the journal's name and declared aims, the existing classifications of similar journals, the publisher's wishes/suggestions and perhaps text mining suggestions. The interdisciplinary expert Content Selection & Advisory Board (CSAB) (Elsevier, 2023a, Elsevier, 2023b) performs this role. It is not clear whether Scopus systematically analyses the classifications given to a journal in subsequent years, although it does have both manual and semi-automatic procedures to deselect journals if they appear to be predatory (Baas et al., 2020).

**Strategy for identifying journal type by publishing practice**: A text-based heuristic was used to match articles to the Scopus classifications to help estimate the field(s) in which the hosting journal publishes. Article topics can be classified through their references, citations, full text, or metadata (Boyack et al., 2011; Klavans & Boyack, 2017). Of these, full text in machine readable format is not widely available for academic journals and both direct citations (Wang & Waltman, 2016) and co-citations are unavailable for most recently published articles. Thus, the two possible options were bibliographic coupling and text metadata. The latter was chosen as a practical step because the article references needed for bibliographic coupling cannot be downloaded from the Scopus Applications Programming Interface (API), and there are too many Scopus articles from 2022 (2.9 million) to manually download all records from the web interface, given the 2000 records per query (with references) download limit. In addition, text similarity methods are relatively transparent in that the main terms causing high similarity scores can be identified. This is more difficult with bibliographic coupling. Thus, journals were assigned to narrow field(s) based on matching text from their metadata (titles and abstracts) with text from the metadata of other journals assigned to each narrow field. More details are given in the next section.

## Data

All journal articles published in 2022 and indexed in Scopus were downloaded in February and March 2023 using the Scopus API in Webometric Analyst, using queries of the form:

SUBJMAIN(1109) AND DOCTYPE(ar) AND SRCTYPE(j), where 1109 is the ASJC (All Science Journal Classification) code for one Scopus narrow field (Elsevier, 2022). Similar queries were submitted for all Scopus narrow fields, with 333 returning at least one match. Each narrow code fits within one of the 27 broad fields with two-digit codes (e.g., 1109 Insect Science is in 11 Agricultural and Biological Sciences). Standardised headings and copyright statements were removed from abstracts with a program (code online: https://doi.org/10.6084/m9.figshare.22183441.v1).

On average, the 333 narrow fields included 193 journals, 22,220 articles and 124.5 articles per journal (Table 2). This includes double counting because each article and journal was whole counted in each narrow field to which they had been assigned.

Table 2. Descriptive statistics for the Scopus-indexed journal articles reported by Scopus narrow field (n=333).

| Statistic | Journals | Articles | Articles per journal |
|---|---|---|---|
| Minimum for any field | 1 | 20 | 11.5 |
| Max for any field | 1,396 | 181,935 | 632.4 |
| Average across all fields | 193 | 22,220 | 124.5 |
| Total across all fields* | 64,213 | 7,399,368 | - |

*Multiply counting multiply classified journals for both articles and journals.

## Journal type by Scopus field(s)

All journals in Scopus are classified into between 1 and 11 ASJC narrow fields and this information was extracted from the article data associated with the journal in the data downloaded with the Scopus API, as above. The narrow field classification scheme has a feature that partly conflicts with the goals of this paper: within each broad field there is a general narrow field, called "all" and a multipurpose narrow field, called "miscellaneous". Thus, for example, within 15 Chemical Engineering there is 1500 Chemical Engineering (all) and 1501 Chemical Engineering (miscellaneous) in addition to seven narrow fields with more specific names, such as 1508 Process Chemistry and Technology. Each of the 27 broad fields except Multidisciplinary contains these two categories that are not narrow fields and a variable number of other narrow fields that from their names appear to be academic fields, in the sense of closely related topics. In total, 52 of the 333 Scopus narrow fields (i.e., 16%) are therefore not narrow academic fields. These were retained for the analysis to allow a complete dataset but are singled out for their special features in discussions of the results.

Journals were categorised by apparent field orientation, according to the Scopus scheme. A journal is *specialist* if it has a single narrow field classification in Scopus. For example, Journal of Librarianship and Information Science has the single classification 3309 Library and Information Sciences, and so is regarded as a specialist journal here. In contrast, a journal is called *cross-field* if it has multiple narrow field classifications in Scopus. Since Journal of Information Science is in two narrow fields (3309 Library and Information Sciences; 1709 Information Systems), it is cross-field here. This is an oversimplification for many reasons. First, a "cross-field" journal could be cross-field in the sense of deliberately including selected largely unrelated fields, cross-field in the sense of a generalist journal spanning multiple related fields, or inclusive/general in the sense of spanning many fields or all science (e.g., Science, Nature). Moreover, a journal might span a narrow field but have multiple classifications because the field crosses the border of two Scopus narrow fields, perhaps because it evolved after the Scopus classification scheme developed. Finally, academic fields are subjective and field members might well disagree on its scope is or what constitutes an academic field. For example, some scientometricians might consider scientometrics to be a field in its own right, whereas others might consider it to be part of library and information science or even perhaps science and technology studies or research evaluation. Despite these limitations, the definitions here at least transparently differentiate between journals that are more likely to be specialist and those that are more likely to be cross-field, despite containing simplifications and probably also some errors.

The cross-field category includes open access mega-journals that have a soundness-only reviewing model, publish large numbers of articles and have a broad scope (see: Wakeling et al., 2019). This broad scope is presumably reflected in multiple relevant Scopus narrow field(s) (or the Multidisciplinary narrow (and broad) field(s)) and the soundness-only reviewing model seems unlikely to affect the text analysis method used here, so they are not treated separately.

## *Journal type by publishing practice*

**Term extraction**: The publishing practice of a journal was identified through the terms in its title and abstract, comparing them to the average terms used in each Scopus narrow field using the cosine similarity measure as follows. Keywords were not used since not all articles have them and some journals use controlled vocabularies, differentiating them from others. Here a "term" means 1-3 consecutive words within a sentence inside the title or abstract. Adding short phrases of 2 or 3 words is helpful because of the number of academic terms that are multiword expressions. Following this procedure, the title "Abbreviations and short titles" would translate into the following nine terms: abbreviations, and, short, titles, "abbreviations and", "and short", "short titles", "abbreviations and short", and "and short titles". All words within a stop word list of 120 common words were removed since these add little meaning (see Appendix). This included common general words (e.g., "a") and common stylistic terms that are not directly related to the topic of an article, such as "herein", "paper", and "article". The list included "and", so in the above example, the six terms extracted from "Abbreviations and short titles" would be: abbreviations, short, titles, "abbreviations short", "short titles", and "abbreviations short titles".

**Journal TF-IDF vectors**: For each journal, a list of terms occurring in its articles (titles and abstracts) was extracted, recording the number of articles containing each term. This Term Frequency (TF) is an article (title and abstract) count rather than a term count, so multiple occurrences of the same term within an article would not increase the journal term count. The TF was then multiplied by the Inverse Document Frequency (IDF) to get a TF-IDF vector for each journal (Manning et al., 2008). The IDF of a term $i$ is $log(N/n_i)$, where N is the overall number of journals and $n_i$ is the number of journals containing term $i$. The TF-IDF score of a term in a journal is therefore high if the term occurs in many of that journal's articles but in few other journals' articles. Conversely, a term's TF-IDF score is low if it is either rare in the journal or occurs in most other journals. The TF-IDF formula is common in information retrieval and for document clustering (Whissell & Clarke, 2011) and so is a standard choice here for creating vectors to represent the text content of a set of documents.

**Field TF-IDF vectors**: The above procedure was repeated for each Scopus narrow field using the same IDF calculation but with each TF being the weighted sum of the TFs of the articles in the journals within the field, with the weight $1/f_j$, where $f_j$ is the number of fields that journal $j$ is classified into. Thus, the TF of term $i$ in field $F$ can be calculated by

adding $1/f_j$ for every article $a$ containing the term, where the article is in a journal in the field:

$$TF_{Fi} = \sum_{j \in F} \sum_{a \in j | i \in a} 1/f_j$$

The TF-IDF vector for a field thus represents the use of terms for articles in journals classified within the fields, weighted for the extent to which the journal is in the field. For example, a journal only classified within a given field would have its terms weighted 11 times higher than the terms of a journal that was also classified in 10 other fields.

**Journal-field similarity calculations**: To test how close a journal's publishing practice is to a field, the cosine similarity between the journal and the field was calculated using the vectors as described above. If a journal was classified as being within the field by Scopus, then the TF-IDF calculations for the field were recalculated without that journal to avoid giving the journal a misleadingly high similarity score for the fields containing it.

**Journal publishing practice classifications**: All journals were classified for publishing practice in terms of articles, relative to each Scopus narrow field. Scope classifications ignore journal aims/Scopus classifications and use only title and abstract cosine similarity information.

- A journal is classified as having a **specialist publishing practice in a Scopus narrow field** if its cosine similarity is highest for that field and its similarity with all other fields is at most 75% as high. The choice of 75% is arbitrary here after inspection of the data. It seems intuitively high enough to ensure that a journal has a primary publishing practice within the field.
- A journal has a **specialist and cross-field publishing practice in a Scopus narrow field** if its cosine similarity is highest for that field and it has a similarity with at least one other field that is above 75% as high. These journals have a similar publishing practice on at least two fields with none clearly dominating.
- A journal has an **cross-field publishing practice with a narrow field** if that field is amongst the Scopus narrow fields that the journal is 2$^{nd}$ to 5$^{th}$ most similar to (i.e., a top 5 similar field, but not the top field). The second category is separated out from the first and third, despite not being in the research questions, for journals with publishing practices that are neither clearly specialist nor clearly cross-field.
- The remaining journals have an **unrelated research publishing practice relative to a Scopus narrow field**. Thus, each journal will be classified once in either of the first two categories, four times in the third category and the remaining times in the last category. The choice of five fields for cross-field is again relatively arbitrary here. Most journals have three or fewer classifications, so allowing five should ensure that few journals are forced to have scopes within their main fields.

**Limitations**: There are many limitations of the term comparison approach, including polysemy and homonymy but most seem likely to be noise in the data and therefore cancelled out in the cosine calculations. Geographic terms seem more likely to cause systematic disruption to the cosine calculation, however. For example, Spanish-language journals of law and politics might mention the same countries and major cities in different contexts. Since such words would be rare, they could have a large influence on the results. Scopus requires journals to publish English titles and abstracts and normally reports these through the API without the original translations. In a few cases, non-English abstracts are indexed, however, apparently in error. For example, there was one such abstract out of the

18859 in field 3309 (for the IC Revista Cientifica de Informacion y Comunicacion paper: "You will never make it, you are too pretty:" Voices of women researchers in communication).

**Alternatives**: Marginally more accurate cosines might have been achieved if the SciBERT (Beltagy et al., 2019) language model had been used to detect the senses of the terms in the abstracts rather than using the ngram approach. This was not used because a transparent solution was needed to analyse the reasons for the results. The BM25 formula has also been shown to be better than TF-IDF (Boyack et al., 2011; Whissell & Clarke, 2011) for some single term text classification tasks but it is asymmetric, more difficult to interpret, and the evidence is not strong enough yet to make it the default for document similarity measurement.

## Research questions

To address RQ1, field differences in the extent to which specialist Scopus journals publish articles that match their aims are identified by examining the Scopus narrow fields with the most large journals (so the data is most reliable), comparing between them for the extent to which specialist journals fall into the four publishing practice types

RQ2 (do cross-field Scopus journals publish articles that match their aims) was addressed as for RQ1 except for cross-field journals.

RQ3 (do general Scopus journals ever have more specialist publishing practices?) was addressed by calculating the similarity of each general journal with the Multidisciplinary field and comparing this to the highest similarity score with any field. Only large journals (over 100 articles in 2022) are reported to give more reliable data.

RQ4 (are there field differences in the type [specialist, specialist and cross-field] of out-of-field journals) was addressed as above except for out-of-field journals.

RQ5 (Why do some journals mainly publish articles not matching their Scopus classifications), was addressed by manually checking selected journals with publishing practices not matching their Scopus narrow field(s). Possible reasons for the apparent mismatch were sought qualitatively by examining (i) the names of the journals, (ii) their article titles, and (iii) terms with the highest TF-IDF weight for the journal compared to the field (somewhat like: Zhang et al., 2016). The first two may point to obvious answers whereas the last would give the most direct evidence of the reason for the similarity by pointing to relatively distinctive article terms/concepts. In theory, this could point to obvious classification errors, spurious reasons (e.g., a Journal of Victorian Studies large special issue on "the heart as a metaphor" causing it to match field 2705 Cardiology and Cardiovascular Medicine), field overlaps, or journal scope changes. This analysis focused exclusively on out-of-field journals in Scopus narrow fields for which there was at least one out-of-field specialist publishing practice journal (i.e., a journal not in a Scopus narrow field but with a specialist publishing practice in that field, as defined above). The reason for this is that such fields seemed most likely to reveal systematic rather than ad-hoc causes, such as special issues.

# Results

The results are reported for 1-3 grams, but similar results were obtained for 1grams (i.e., single words).

## RQ1: Specialist (and Multidisciplinary) journal Scopus narrow field vs. publishing practice

There are almost the maximum possible field differences in the extent to which specialist journals (i.e., those with a single Scopus AJSC narrow field classification) have a matching specialist publishing practice (Fig. 1). At one extreme, 95% (20/21) of Dermatology specialist journals have the same specialist publishing practice (i.e., on Dermatology), whereas only 4% (2/45) of Mathematics (all) journals do. It is noticeable that there are many general "(all)" narrow fields in the list, presumably containing journals that are general in scope but categorised as specialist with the simplistic definition in the current paper. The same is clearly true for the Multidisciplinary field. These fields contain few purely specialist publishing practice journals, if any, and varying amounts of the other types of journals. Ignoring these interdisciplinary narrow fields, however, the specialist journals in ten narrow fields are at least 80% (purely) specialist in practice, compared to those in five narrow fields that are under 5% to 67% specialist in publishing practice.

The difference between journals with a specialist publishing practice and those with a specialist and cross-field publishing practice is partly a function of the Scopus classification scheme and the extent to which topics are distinct from others. For example, it seems that Dermatology may be a relatively distinct research topic, but perhaps Mechanical Engineering has overlaps with other narrow fields.

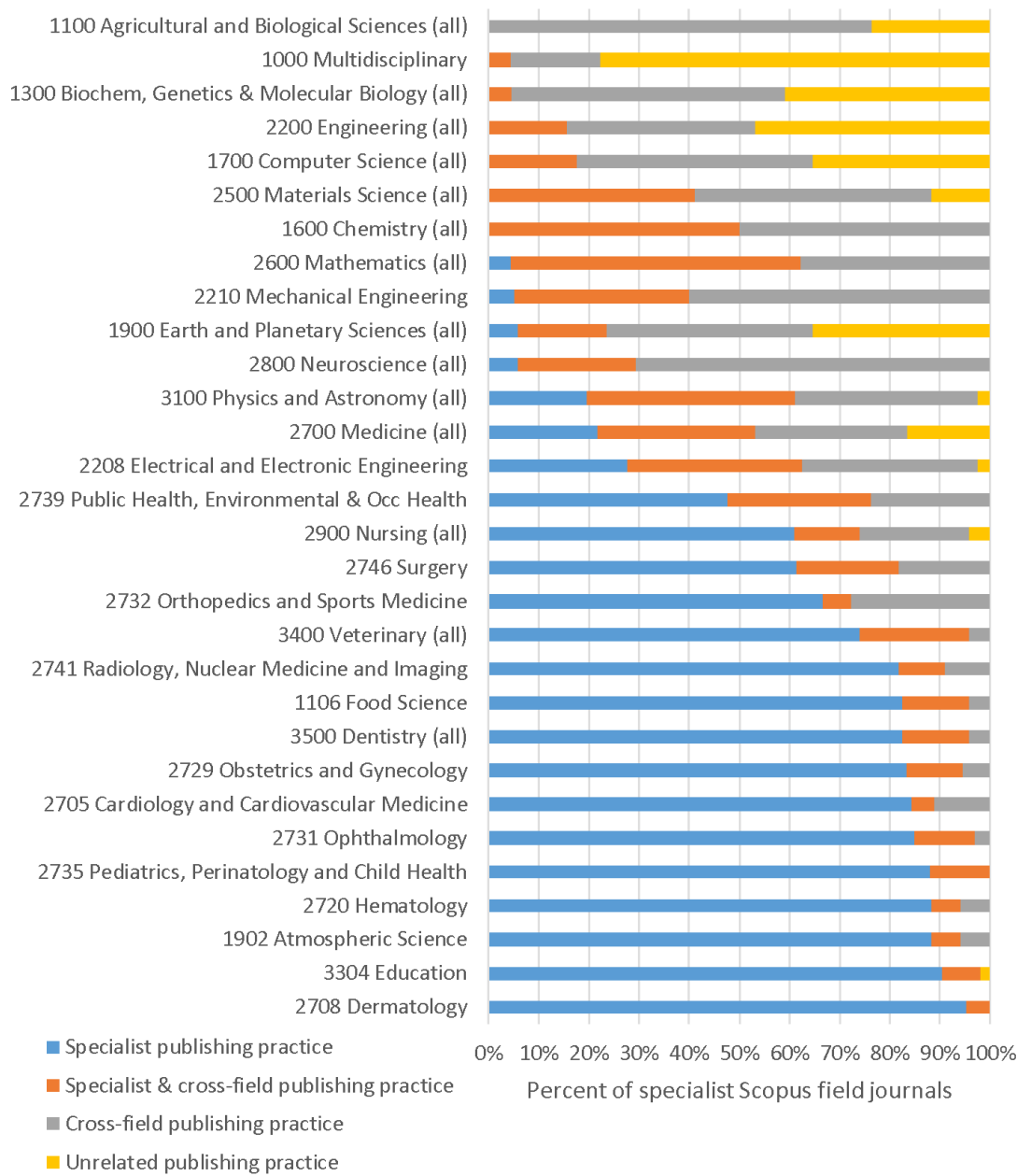

Fig 1. The distribution of journals by publishing practice within Scopus fields for specialist Scopus narrow fields journals (i.e., categorised solely within the named field by Scopus) and the general journals in the Multidisciplinary category. Qualification: The 30 Scopus narrow fields with at least 17 specialist journals with at least 100 articles each in 2022.

## RQ2: Cross-field journal Scopus narrow fields vs. publishing practice

There are moderate field differences in the extent to which cross-field journals (i.e., those with multiple ASJC Scopus narrow field codes), are (relevant) specialist or cross-field in practice (Fig. 2). The proportion of journals with an unrelated research practice varies from 9% (14/154) (Surgery) to 62% (63/101) (Industrial and Manufacturing Engineering). Unsurprisingly, relatively few of these cross-field journals in all fields have a specialist practice, and this occurs the most (40%: 39/98) for Education.

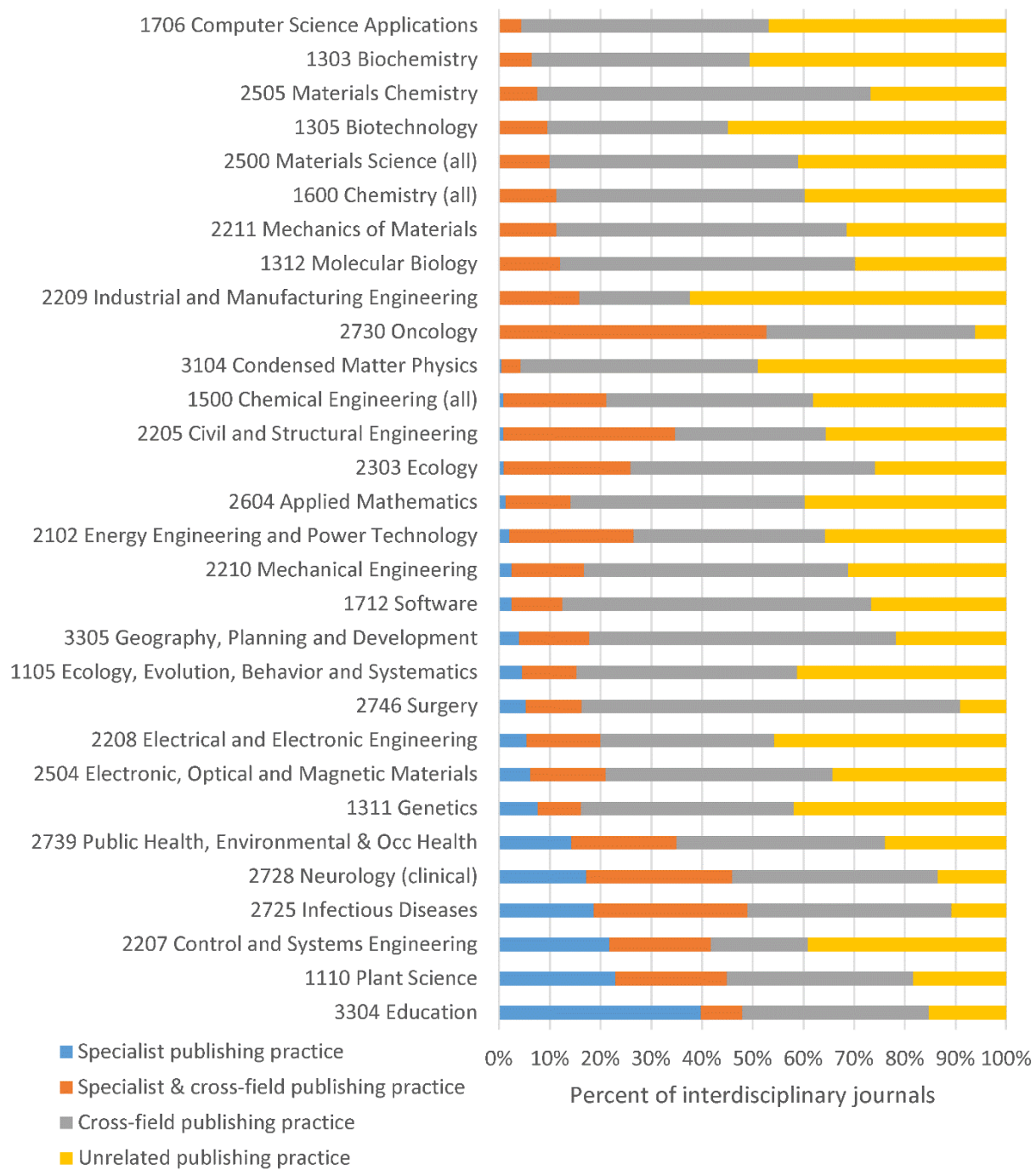

Fig 2. The distribution of journals by publishing practice within Scopus fields for cross-field Scopus narrow field journals (i.e., categorised within the named field by Scopus and at least one other named field). Qualification: The 30 Scopus narrow fields with at least 98 cross-field journals with at least 100 articles each in 2022.

### RQ3: General journals vs. publishing practice

The "specialist" journals in the Scopus Multidisciplinary category (Fig. 1) are ostensibly general journals since they are solely in this general category. There are also some ostensibly partly general journals that have the category 1000 and some other categories (e.g., *Foundations of Science* has the second category, 1207 History and Philosophy of Science). Nevertheless, the ostensibly general journals seem to vary in degree of disciplinarity from the two fully multidisciplinary Science and Heliyon to the many more specialist journals like Journal of Hunan University Natural Sciences (Hunan Daxue Xuebao), which seems to specialise in Civil and Structural Engineering (Table 3) and the UK science magazine, New Scientist, which seems to have a particular interest in Astronomy and Astrophysics. Table 3 also exposes the potential for error in the classification system used, however, because Nature's main area is biology rather than physics or astrophysics (e.g., the top 20 keywords in Nature 2022 according to Scopus are all biological). The cause of the error for Nature is that the Scopus field 3100 Physics & Astronomy (all) is mainly populated with biology due to the presence of Nature Communications in this category. This can be checked by submitting the query SUBJMAIN(3100) AND DOCTYPE(ar) AND SRCTYPE (j) AND PUBYEAR IS 2022 to Scopus and noticing that the top keywords are all biological.

Table 3. Journals with the sole category of 1000 Multidisciplinary in Scopus, with at least 100 articles in 2022. The table reports the journal cosine similarity with the 1000 Multidisciplinary field (excluding the journal itself from the field) as a proportion of the highest field cosine similarity, and the field that it has the highest cosine similarity with.

| Journal | Articles | Sim. w. 1000 | Closest field |
|---|---|---|---|
| Heliyon | 3531 | 1 | 1000 Multidisciplinary |
| Science | 1150 | 1 | 1000 Multidisciplinary |
| Brazilian Archives of Biology & Technology | 147 | 0.612 | 1102 Agronomy & Crop Science |
| Royal Society Open Science | 530 | 0.856 | 1105 Ecology, Evolution, Behavior |
| Sains Malaysiana | 320 | 0.721 | 1106 Food Science |
| Songklanakarin J Science & Technology | 164 | 0.854 | 1106 Food Science |
| Philippine J Science | 214 | 0.617 | 1110 Plant Science |
| Scientific reports | 21849 | 0.934 | 1303 Biochemistry |
| J King Saud University - Science | 587 | 0.689 | 1303 Biochemistry |
| Comptes Rendus de L'Academie Bulgare des Sciences | 214 | 0.965 | 1303 Biochemistry |
| ScienceAsia | 119 | 0.637 | 1303 Biochemistry |
| PNAS | 3461 | 0.737 | 1312 Molecular Biology |
| Science Advances | 2159 | 0.885 | 1312 Molecular Biology |
| iScience | 1850 | 0.725 | 1312 Molecular Biology |
| J Advanced Research | 208 | 0.619 | 1312 Molecular Biology |
| International J Advanced & Applied Sciences | 227 | 0.533 | 1408 Strategy & Management |
| Emerging Science Journal | 129 | 0.433 | 1408 Strategy & Management |
| Trends in Sciences | 354 | 0.725 | 1500 Chemical Engineering (all) |
| J Advanced Research in Applied Sciences & Eng Tech | 125 | 0.373 | 1507 Fluid Flow & Transfer Proc |
| Chinese Science Bulletin | 299 | 0.674 | 1600 Chemistry (all) |
| J Shanghai Jiaotong University (Science) | 202 | 0.387 | 1706 Computer Science Apps |
| Acta Scientiarum Natralium Universitatis Sunyatseni | 114 | 0.674 | 1900 Earth & Planetary Sci (all) |
| Acta Scientiarum Naturalium Universitatis Pekinensis | 119 | 0.363 | 1907 Geology |
| J Xi'an Shiyou University, Natural Sciences Edition | 107 | 0.205 | 1909 Geotechnical Eng |
| Arabian Journal for Science & Engineering | 1415 | 0.421 | 2200 Engineering (all) |
| J Scientific & Industrial Research | 130 | 0.547 | 2200 Engineering (all) |
| J Hunan University Natural Sciences | 347 | 0.267 | 2205 Civil & Structural Eng |
| J Southwest Jiaotong University | 294 | 0.43 | 2205 Civil & Structural Eng |
| J Tongji University | 200 | 0.286 | 2205 Civil & Structural Eng |

| J Tianjin University Science & Technology | 144 | 0.322 | |
| Sadhana - Academy Proceedings in Eng Sciences | 282 | 0.323 | 2209 Industrial & Man Eng |
| J Jilin University (Engineering & Technology Ed) | 334 | 0.327 | 2210 Mechanical Engineering |
| J Shanghai Jiaotong University | 174 | 0.296 | |
| J Taiyuan University of Technology | 135 | 0.421 | |
| Science Progress | 121 | 0.788 | |
| Anais da Academia Brasileira de Ciencias | 435 | 0.623 | 2303 Ecology |
| Current Science | 360 | 0.621 | |
| Scientific African | 313 | 0.775 | 2310 Pollution |
| Fundamental Research | 245 | 0.537 | 2500 Materials Science (all) |
| Science Bulletin | 245 | 0.394 | |
| Research | 173 | 0.502 | |
| National Science Review | 158 | 0.583 | |
| PLoS ONE | 15103 | 0.72 | 2739 Public, Env & Occ Health |
| Nature | 1301 | 0.909 | 3100 Physics & Astronomy (all) |
| New Scientist | 175 | 0.187 | 3103 Astronomy & Astrophysics |

## RQ4: Out-of-field journals (publishing outside their Scopus narrow fields) vs. degree of publishing specialism

Many journals were found with publishing practices outside their Scopus narrow field(s), but the exact number is uninformative because this depends on the size of each narrow Scopus field. The shares by Scopus classified fields are available in the previous two diagrams, and this section focuses instead on the field of publishing practice.

There are substantial field differences in the extent to which out-of-field journals with a publishing practice relevant to a field have a specialist publishing practice in the field, for the fields with the most journals of either type (Fig. 3). In particular, 50% (16/32) of Fluid Flow and Transfer Processes journals with a publishing practice relevant to this field but not classified in it by Scopus have a specialist publishing practice in it, in contrast to 0% for 18 fields.

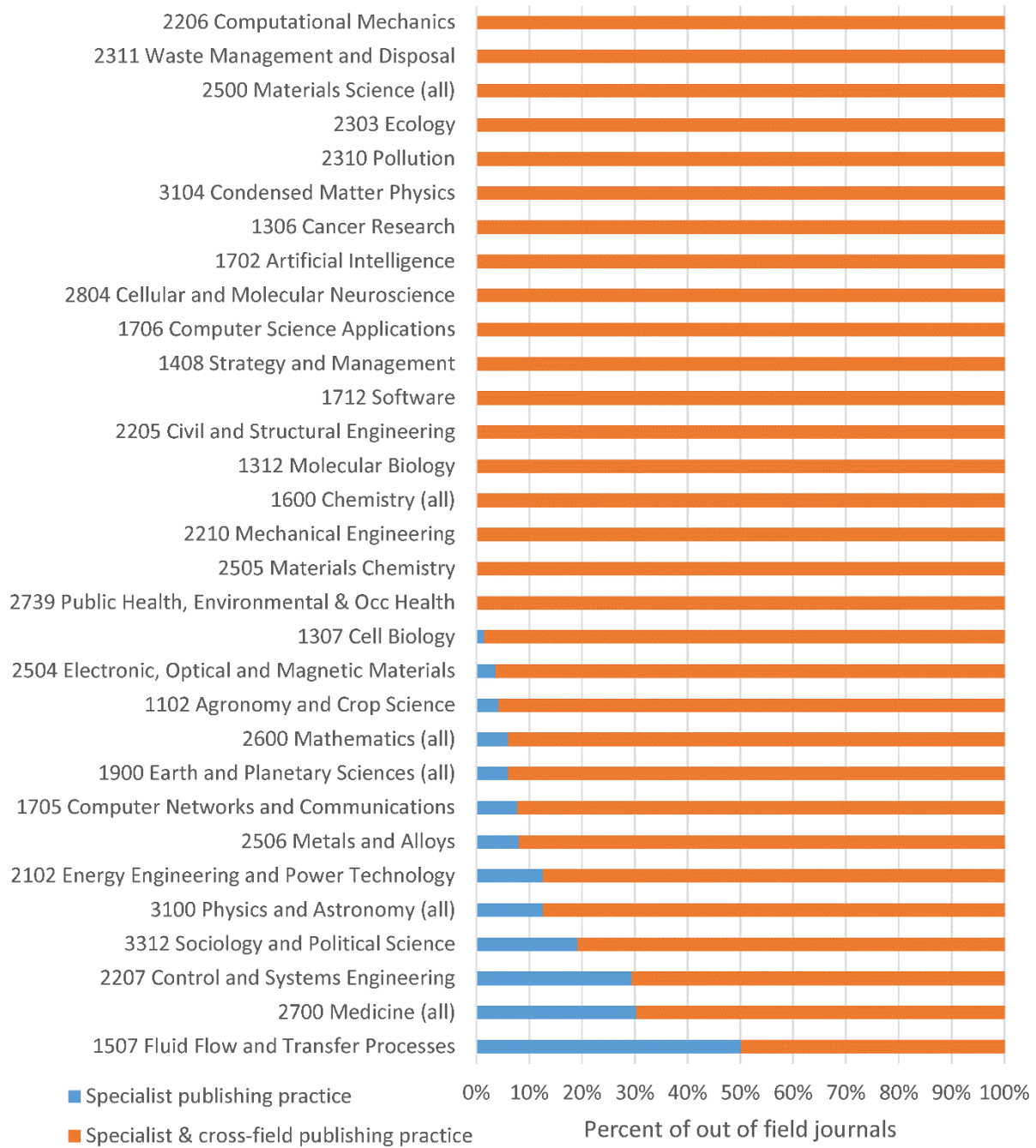

Fig 3. The distribution of journals by publishing practice within Scopus fields for out-of-field Scopus journals (i.e., not categorised within the named field by Scopus but with a publishing practice most similar to the named field). Qualification: The 30 Scopus fields with at least 16 journals containing at least 100 articles each in 2022 in the two categories combined.

## RQ5 Why do some journals have a publishing practice not matching their Scopus narrow field(s)

All out-of-field journals for the 13 fields in Figure 3 with at least one specialist publishing practice journal were examined qualitatively (title, and terms with highest TF-IDF with the relevant field). This set was chosen as the most relevant fields because out-of-field journals are forced to exist by the definition used (i.e., derived from the top five publishing fields per journal, so journals with less than five Scopus categories are forced to be out of field journals for some fields), but no journal is forced to have a specialist publishing practice in a field that is outside of its Scopus categories. The analysis covers both types of journals, however, to give a wider picture.

The results are discussed in detail for one field, and then the results are summarised for the other 12 fields, with details available in the online supplementary materials. For Fluid Flow and Transfer Processes (within broad field Chemical Engineering), the 36 out-of-field journals (with at least 100 articles in 2022) included three that had **appropriately more general ASJC codes** (Multidisciplinary and Chemical Engineering) and the rest had **related specialisms** (Table 4). The underlying cause in all cases seemed to be a common interest in either fluid flow or heat transfer (including through fluid flows) in field 1507 and as a core theme in many of the out-of-field journals. These journals covered a wide range of different fields, but the top TF-IDF terms tended to relate to heat transfer (e.g., heat_transfer, nusselt_number). The journals were mostly from areas of engineering, physics, and applied mathematics, which are clearly relevant to heat transfer problems. In contrast, Journal of Visualization seems irrelevant from its title, but it had published many papers about fluid flow in 2022 (e.g., 9 with the keyword Flow Visualization), and before (Flow Visualization is its all-time second most popular keyword, according to Scopus, behind only Visualization). Microgravity Science and Technology might also seem irrelevant but its all-time top eight keywords, according to Scopus, include Heat Convection, Liquids, Capillary Flow, and Heat Transfer. Moreover, its top TF-IDF term with 1507 is heat_transfer.

Table 4. Out-of-field journals (32) with at least 100 articles in 2022 for 1507 Fluid Flow and Transfer Processes.

| Journal | Articles | Fields | Reason/overlap |
|---|---|---|---|
| Journal of Advanced Research in Applied Sciences and Engineering Technology | 125 | 1000 | More general |
| International Communications in Heat and Mass Transfer | 666 | 1500, 3104, 3107 | More general |
| Chemical and Petroleum Engineering | 140 | 1500, 1906, 2102, 2103 | More general journal |
| Journal of Thermal Analysis and Calorimetry | 1173 | 1606, 3104 | Thermal energy |
| European Physical Journal: Special Topics | 361 | 1606, 2500, 3100 | Related specialism |
| Engineering Applications of Computational Fluid Mechanics | 120 | 1700, 2611 | Fluid Flow |
| International Journal of Modern Physics C | 229 | 1703, 1706, 2610, 3100, 3109 | Related specialism |
| International Journal of Numerical Methods for Heat and Fluid Flow | 192 | 1706, 2210, 2211, 2604 | Thermal energy |
| Flow Measurement and Instrumentation | 128 | 1706, 2208, 2611, 3105 | Fluid Flow |
| International Journal of Modelling and Simulation | 135 | 1708, 2200, 2208, 2209, 2211, 2600, 2611 | Related specialism |
| Applied Thermal Engineering | 1351 | 2102, 2209 | Thermal energy |
| Proceedings of the Institution of Mechanical Engineers, Part A: Journal of Power and Energy | 151 | 2102, 2210 | Thermal energy |
| International Journal of Ambient Energy | 805 | 2105, 2215 | Thermal energy |

| Journal | | | |
|---|---|---|---|
| Thermal Science | 486 | 2105 | Thermal energy |
| Alexandria Engineering Journal | 933 | 2200 | Related specialism |
| Waves in Random and Complex Media | 916 | 2200, 3100 | Related specialism |
| International Journal of Thermal Sciences | 519 | 2200, 3104 | Thermal energy |
| Journal of Engineering Physics and Thermophysics | 199 | 2200, 3104 | Thermal energy |
| Microgravity Science and Technology | 101 | 2200, 2604, 2611, 3100 | Related specialism |
| Journal of Applied and Computational Mechanics | 117 | 2206, 2210 | Related specialism |
| Experimental Heat Transfer | 115 | 2207, 2208, 3105 | Thermal energy |
| Journal of Visualization | 112 | 2208, 3104 | Related specialism |
| Kung Cheng Je Wu Li Hsueh Pao/Journal of Engineering Thermophysics | 423 | 2210, 2500, 3104 | Related specialism |
| Reneng Dongli Gongcheng/Journal of Engineering for Thermal Energy and Power | 290 | 2210, 3104 | Thermal energy |
| International Journal of Refrigeration | 273 | 2210, 2215 | Thermal energy |
| Journal of Fluids Engineering, Transactions of the ASME | 194 | 2210 | Fluid Flow |
| Journal of Heat Transfer | 157 | 2210, 2211, 2500, 3104 | Thermal energy |
| Journal of Applied Fluid Mechanics | 149 | 2210, 2211, 3104 | Fluid Flow |
| Fluid Dynamics and Materials Processing | 154 | 2500 | Fluid Flow |
| Numerical Heat Transfer; Part A: Applications | 125 | 2612, 3104 | Thermal energy |
| Journal of Thermal Science | 187 | 3104 | Thermal energy |

Summarising the results for out-of-field journals in all fields, the following are the main apparent causes.

- The **journal has a narrower classification than the field** but either its narrow field contains diverse topics, or the more general field publishes relatively many articles in the specialist area of the journal.
- The **journal has a broader classification than the field** (either 1000 Multidisciplinary or the field's parent field with "(all)" in its name) but tends to be more specialist than its apparent aims suggest. This seems particularly likely to occur for journals in "(misc)" narrow fields that might not match their home narrow field well.
- The **journal covers a related topic to the field**. For example, Visualisation relates to fluid flow because of the extensive development of fluid flow visualisations. The out-of-field effect can occur if the journal publishes on very different topics, so does not fit any field well. This can be exacerbated by the journal being classified into a general field (e.g., applied computing rather than visualisation).
- The **journal's topic is cross-field with extensively overlapping research in multiple fields**. This seems to be the case for fluid flow in the context of heat transfer, for example, as well as for the field Control and Systems Engineering connecting to robotics and systems research. The same is true can be true for journals relating to an overlapping aspect of two fields, such as politics journals in the fields Sociology and Political Science or Political Science and International Relations.

- The **journal scope is cross-field and fits no field well**. For example, npj Computational Materials has four non-Physics classifications but is an out-of-field journal for the Physics and Astronomy (all) narrow field.
- The **Scopus classification can be wrong**. For example, Rawal Medical Journal is classified as Nursing (all) but Medicine (all) fits its title, aims, and articles better. The Philosophical Magazine classification also seems to be wrong or outdated since its aims mention physics and materials, but its five Scopus narrow fields exclude Metals and Alloys, which it is an out-of-field journal for.
- The **journal's scope does not match its aims from an outsider perspective**. This rarely happened but Organization Studies was an out-of-field journal for Sociology and Political Science through discussions of pollical issues like capitalism, neoliberalism, democracy, and resistance.

There was also weak evidence that overlapping methods from different fields helped to bring the journals of the two fields closer in terms of TF-IDF scores. This might have been the explanation for some toxicology out-of-field journals for Cell Biology. Mouse models were used in both cases. Nevertheless, this case could also be explained by overlapping topics (e.g., toxic chemicals causing cell death). Other than this, there was no evidence of out-of-field journals being unrelated to their field but only matching because of non-semantic reasons like key term polysemy.

## Discussion

The results are limited by the database chosen, the year of analysis, and the many methodological heuristics and assumptions, so should not be interpreted as conclusive. In particular, the definitions of journal types are moderated by the Scopus classification scheme, as are the definitions of a journal's publishing practice, as well as containing an arbitrary threshold (0.75). Another important limitation is that the results apply exclusively to relatively large journals with at least 100 articles in 2022. Many arts and humanities journals seem to be much smaller, so they are underrepresented.

The results confirm, with different methods and updated data, previous findings that there can be journal classification errors in Scopus and misleading journal scopes (Wang & Waltman, 2016). The same was true for the Web of Science (Leydesdorff & Bornmann, 2016; Wang & Waltman, 2016).

In answer to RQ1, there are substantial field differences in the extent to which apparently **specialist** journals publish articles that match their Scopus classification, from Dermatology (almost all specialist journals have a publishing practice in the area) to Mechanical Engineering (almost no specialist journals with a strong specialist publishing practice in the area and 60% having a cross-field publishing practice). There does not appear to be a systematic pattern in the results in the sense of broad disciplinary areas. Instead, the results seem to reflect the extent to which research interests for a field are distinct to those of other fields. As the case of fluid flow research discussed above suggests, field overlapping might occur due to societal issues that cause different fields to work on shared societally relevant problems, such as energy efficiency. Thus, for example, fluid flows might be relevant for visualisation because understanding fluid flow in energy transfer is an important societal issue.

In answer to RQ2, there are also substantial field differences in the extent to which **cross-field** journals publish articles that match their Scopus classifications. For example, whilst 93% of cross-field journals with aims encompassing Oncology have a related publishing practice, the same is true for only 38% of Industrial and Manufacturing Engineering journals. This seems likely to be again related to the extent to which fields are distinct, with Oncology perhaps being relatively specific (although relating to other fields) compared to Industrial and Manufacturing Engineering.

In answer to RQ3, ostensibly **general** journals often match non-general Scopus narrow fields much better than Multidisciplinary, confirming that ostensibly general journals can still have specialisms. This is an almost tautological conclusion because the text comparison method relies on the Multidisciplinary category being fully general, whereas the results show that it is not. Logically, however, it is not possible for the Multidisciplinary category being fully general with these results, although it is logically possible (and likely) that no journal is fully general, despite two large general journals being more similar to the Multidisciplinary category than to any other.

In answer to RQ4, there are substantial field differences between fields in the extent to which **out-of-field** journals are specialist, rather than specialist and multidisciplinary. It is perhaps surprising that any fields have a substantial proportion of specialist out-of-field journals (i.e., journals not classified in the field by Scopus but with articles and titles textually most similar to the field).

In answer to RQ5, there are multiple reasons why some journals mainly publish articles not matching their Scopus classifications. First, the aim statement can be misleading or out of date (Wang & Waltman, 2016), confusing the Scopus classification team. Second, the journal's publishing practice may appear not to match its aims for outsiders because they would not expect its theory (e.g., politics in organisation studies), methods or topic (e.g., fluid flow in visualization research). Third, the journal may be cross-field or general but mainly publishes in one field, perhaps for historical reasons or because it fills a publishing gap in that field. Fourth, a journal's topic may have started to overlap substantially with that of other fields due to societal needs increasing research demand for it, such as heat transfer.

The problems discussed above may be part of the reason why article-level automatic classifications of journals can be superior to journal-based classifications of journals (Klavans & Boyack, 2017).

## Conclusion

The results suggest that specialist, cross-field and general journals (as categorised from their Scopus classifications) do not always publish articles that match their Scopus classifications and probably also their declared aims. In extreme cases, apparently unrelated journals can publish substantially in a field and there are few genuinely specialist journals (e.g., Mechanical Engineering). Moreover, there are apparently non-systematic differences between fields in the extent to which journal contents match their Scopus classifications, so there is no general rule about the types of fields in which journal classifications are reliable.

Because of these mismatches between journal classifications and publishing practices, authors, readers, and research evaluators should be careful to not make assumptions about the scope or content of a journal from its Scopus classification or title and aims statement. When conducting a literature search for a field with unknown journals, care should also be taken to avoid ruling out articles in apparently unrelated journals. Similarly, scholars browsing journals or systematically checking by journal should ensure that their initial identification of relevant journals includes methods of identifying ostensibly irrelevant journals that nevertheless publish relevant research. For example, a keyword search of Scopus, the Web of Science or Dimensions, followed by an examination of the lists of matching journals may reveal previously unknown titles. Similarly, authors seeking journals to publish in should be careful to ensure that their submission matches the publishing practice of their intended journal and not just its Scopus classification, aims or title.

Perhaps the most serious implication for research evaluation, echoing in more detail the finding of a previous study (Wang & Waltman, 2016), is that the journal categorisation process in Scopus can be unreliable and this may undermine the accuracy of citation indicator calculations that depend on this. These include field normalised citation scores and citation-based league tables of journals. The problem cannot be resolved by focusing on specialist journals only since these may have publishing practices that differ from their aims.

**Declarations**

*Competing interests*: The first author is a member of the Distinguished Reviewers Board of Scientometrics.

*Funding*: No funding was received for conducting this study.

## Appendix: Stop words used

a, about, across, after, all, also, among, an, analysis, and, any, are, article, as, at, be, because, been, being, between, both, but, by, can, could, despite, did, do, each, first, five, for, found, four, from, had, has, have, having, hence, here, herein, high, how, however, if, in, into, investigate, investigated, is, it, like, many, may, method, more, moreover, most, much, namely, no, not, of, on, one, only, or, other, our, out, over, paper, particularly, present, result, second, several, should, show, showed, since, six, so, some, study, such, than, that, the, their, then, there, therefore, these, they, this, those, three, through, thus, to, two, upon, used, using, was, we, were, what, when, where, which, while, who, will, with, without, would, yet.